\documentclass[11pt]{article} 
\usepackage{setspace}
\usepackage[letterpaper, total={16.5cm, 22.5cm}]{geometry}   

\makeatother %
\makeatletter
\let\@fnsymbol\@arabic
\makeatother

\usepackage{multirow, bm}
\usepackage{amsmath,amscd,amsfonts, graphics, color}
 \usepackage{amssymb, latexsym, framed, rotating}
  \usepackage{textcomp}
\usepackage{eurosym}
\usepackage{hyperref, apacite} 
 \usepackage{cite}

\usepackage[figure]{hypcap}
 \usepackage{booktabs}
\usepackage[format=plain, indention=0cm]{caption}
\usepackage{longtable}	
\usepackage{tabularx}

\usepackage{pdflscape}
\usepackage{adjustbox}
\usepackage{changepage}

\title{Feasibility of using survey data and semi-variogram kriging to  obtain bespoke indices of neighborhood characteristics: A simulation and a case study}

\author{
Emily Finne and Odile Sauzet\\ Department of Business
Administration and Economics, Bielefeld University, \\Bielefeld, Germany;\\  School of Public Health, Bielefeld University, Bielefeld, Germany.}

\begin{document}

\maketitle
\abstract{\it{Data on neighbourhood characteristics are not typically collected in epidemiological studies. They are however useful in the study of small-area health inequalities and  may be available in social surveys. We propose to use kriging based on semi-variogram models  to predict  values at non-observed locations with the aim of obtaining indicators of neighbourhood characteristics of epidemiological study participants. The spacial  data available for kriging is usually spare at small distance and therefore we perform a simulation study to assess the feasibility and usability of the method as well as a case study using data from the RECORD study. Apart from having enough observed data at small distances to the non-observed locations, a good fitting semi-variogram, a larger range and the absence of nugget effects for the semi-variogram models are factors leading to a higher reliability. }}

\newpage
\section*{Background}

Health of individuals is not just determined by individual characteristics but also by the environment, physical and social, in which one lives\cite{DiezRoux.2016, Kress.2020, Zolitschka.2019} such that an unequal distribution of health determinants may lead to health inequalities \cite{Zolitschka.2022, Egan.2008}. In particular, the spatial distribution of contextual factors such as access to transport, noise but also neighbourhood social structures, play a role in the development of health inequalities\cite{Zolitschka.2022}. In social epidemiological research, there is a wide range of possibilities to empirically assess contextual factors. Noise or air pollution data obtained from official sources at very small-scale, for example,  provide an accurate description of exposure at the location of study participants. Contextual factors can also be measured as individual perception, data which is typically available in social surveys. While these may be poor proxies for objective measures, they are determinants of health in their own rights. Concerning neighbourhood social structures there may be no other more objective measures \cite{Zolitschka.2019}.  

In this research context, indicators of small-area (e.g. neighbourhood) characteristics are used to study  contextual effects on health inequalities. An example of such indicators are   indices of multiple deprivation (IMDs),  tools to categorise the socio-economic structure of the place of residence of study participants. These are usually obtained from official statistics or surveys and provide  aggregated indicators at the level of administrative units.  This means that indices are based on dimensions which are the same for everyone living in a given unit. A example of such indices available for analysis in social-epidemiology  is the one developed for all nations of the U.K. \cite{Abel.2016} or the German Index of Multiple Deprivation (GIMD) \cite{Maier.2018}. IMDs are  usually categorised into quintiles \cite{Abel.2016} and  linked to observations based on their area of residence. A reason for this is to avoid the  assumption of  a linear association between outcome and deprivation. 

Various methods have been developed to increase the usability of IMDs and widen the source of data used. One example is to use regression models to adjust IMDs obtained separately for the  the four U.K. nations based on different data and statistical units \cite{Abel.2016}. Regression models are also used to "link" indicators estimated from surveys valid at a larger scale using small-area shrinkage estimation methods \cite{G.Datta.2012}.  More generally, Gotway and Young \cite {Gotway.2002, Gotway.2007} proposed a framework for linking geographically aggregated data from different sources for the purpose of spatial prediction providing measures of uncertainty. Spatial microsimulation is also a method linking various data sources to create synthetic datasets  by combining, for example,  individual survey data with small area aggregated data (e.g. census data). These datsets are then used in simulation studies  \cite{Smith.2021}. This method  had been used to study various questions relative to contextual health inequalities \cite{Campbell.2016, Ballas.2006, Wu.2022}. One particular limitation of all these procedures for their application to the  study of small-area health inequalities  is their reliance on administrative units at which data is aggregated. This limitation has been highlighted in the neighbourhood effect literature \cite{van2011neighbourhood} 

Moreover IMDs are not necessarily relevant for all the population concerned or research questions. In particular the small-area determinants of health of vulnerable populations can be different from those of the general population.
Theoretical work on small-area factors relevant for the health of vulnerable populations (e.g., refugees) postulate that the housing context is relevant for health, in particular "immediate surroundings of the accommodation and its’ physical and social boundaries" \cite{Dudek.2022c}. A quantitative analysis confirms that feelings of isolation, of insecurity and limited contact with neighbours are associated with the health of refugees \cite{Dudek.2022b}. These factors, specific to a particular population group, might not be relevant for the majority population. A major point is that the relevance of these neighbourhood characteristics is only  very local and it would not make sense to use them aggregated at a larger scale. Therefore we need alternative methods to obtain  indicators of neighbourhood characteristics  based on bespoke variables and defined at very small spatial scales. Valuable information about neighbourhood characteristics are contained in social survey data in which participants are asked about their perception of green space, noise, quality of construction, or about the social characteristics of their neighbourhood. Aggregated as a score, these variables  provide neighbourhood  metrics based on individual perception.

Geo-spatial modelling methods, in particular modelling semi-variograms, have been successfully used to evaluate non-measured spatial effect on health\cite{Breckenkamp.2021}. This consists of estimating how much an outcome is correlated to the outcome of a neighbour as a function of the distance to the neighbour\cite{Schabenberger.2005}. Information about neighbourhood characteristics at small scale and in particular relative  to the social fabric of a neighbourhood may be available in survey data for which the geo-location of the participants is included. Modelling the spatial distribution of this data provides models which can be used for the prediction of values at given locations via the kriging method\cite{Cressie.1993,Schabenberger.2005}. This would enable to obtain  bespoke indicators of neighbourhood characteristics to be used in the study of small-area health inequality for specific populations at individual location rather than aggregated data at a given spatial unit.

It remains to show whether this approach provides reliable and useful predictions of the true values. A particular source of error is the estimation of the semi-variogram from sparse data at small distances as seen in Sauzet et al. \cite{Sauzet.2021}. Consequently, only a small number of data points is available for the kriging estimates at small distances, given that the weight attributed to data beyond the range of the semi-variogram should be virtually null. Different variables  related to aspects of a common construct like, for example,  social cohesion \cite{Kress.2020} can be correlated at the same location as well as  across different spatial locations. These correlation structures can also be modelled using a form of semi-variogram generalised to several variables: the co-variogram\cite{Gelfand.2010b}. A score  can be constructed as linear combinations of variable values predicted separately from each other or by using a common co-variogram model. 

In this work we investigate the usability of kriging models based on semi-variograms to predict geo-located neighbourhood characteristics  at small scale (less than a kilometre) based on data of the type available in survey data: large sample size but sparse data at small distances. We first recall the definition and estimation of semi-variograms and the method of kriging. We then perform a simulation study to evaluate the reliability of semi-variogram based kriging prediction when data at small-distance is sparse, and next extend the simulation to a comparison of univariate semi-variograms with co-variograms to build indicators based on several variables. Finally, we present a case study based on data from the  RECORD study collected of over 7,000 inhabitants of the Urban Area of Paris \cite{Chaix.2012}.

\section*{Method}
\subsection*{Statistical model}\label{sec:estvario}
We assume a Gaussian random field as random continuous process underlying the distribution of a spatially correlated neighborhood characteristic ${Z}$. More details on what follows can be found in \cite{Schabenberger.2005}. That is, ${Z}$ is assumed to exist in every point $\mathbf{s}$ defined by its coordinates $(x,y)$ of a continous area $D$.  We denote by $Z(\mathbf{s})$ the realisation of the random variable $Z$ at locations $\mathbf{s}$. We also assume second order stationary and isotropy of the random field. That means that the covariance of $Z$ measured at two different locations only depends on the distance (lag) $h$ between the locations and is independent of its direction (rotation-invariant), and the random field has a constant mean: \[E(Z(\mathbf{s})) = \mu; \quad Cov(Z(\mathbf{s}), Z(\mathbf{s}+h) = C(h).\] It follows that the variance of $Z$ is also constant since it corresponds to $C(\mathbf{0})$.

In geostatistics, it is common to express the spatial covariance structure of a continuous process in terms of \emph{semi-variograms}. The semi-variogram contains mainly the same information as the covariance function. It is defined as:
\begin{align}
\gamma(\mathbf{s, h}) 
&= \frac{1}{2} Var\left[Z(\mathbf{s}) - Z(\mathbf{s + h})\right] \notag \\
&= \frac{1}{2} \{Var \left[Z(\mathbf{s})\right] + Var\left[Z(\mathbf{s} + h)\right] - 2Cov\left[Z(\mathbf{s}), Z(\mathbf{s}+h) \right] \} \notag.
\end{align}   
And for second order stationary processes one has:
\begin{align}
\gamma({h}) = \frac{1}{2} \{ 2\sigma^2 - 2 C({h})\} = C(\mathbf{0}) - C(h) \notag
\end{align} 
such that 
$\gamma(\mathbf{0}) = C(\mathbf{0}) - C(\mathbf{0}) = 0$, 
the semi-variogram passes through the origin, and it reaches (approximately) 
$\gamma({h^*}) = \sigma^2$ the partial sill when the distance ${h}$ exceeds the distance ${h^*}$, the so called range, up to which observations are correlated\cite{Schabenberger.2005}.

The empirical semi-variogram is estimated by the Mathéron estimator as 
\begin{align}
\hat{\gamma}({h}) = \frac{1}{2\vert N({h})\vert} \sum_{N({h})}^{} \{Z(\mathbf{s_i}) - Z(\mathbf{s_j}) \}^2 \notag
\end{align}
where $\vert N({h})\vert$ is the number of pairs of data points which have a lag distance of $h$ (or lie in a small interval around).  \\
The Mathéron estimator is unbiased under the above mentioned hypotheses. The graph of $\hat{\gamma}({h})$ against $\Vert {h}\Vert$ is called the \emph{empirical semi-variogram}. 

There are cases where the (empirical) semi-variogram does not pass through the origin because of a) spatial microprocesses whith a more fine-grained resolution as reproduced by the semi-variogram, or b) because of measurement errors. In practice the variables are not completely spatially structured and therefore the correlation of two observations taken at the same location are not necessarily equal. The deviation from the origin at $h =0$ is called \emph{nugget effect} $c_0$, with $c_0 + $partial sill $\sigma_0^{2} = \sigma^2$. \\
An empirical semi-variogram often can be approximated by fitting one of several standard functions. One standard parametric model is the \emph{exponential} semi-variogram which can be fitted as
\[\hat{\gamma}(h) = \hat{c}_0 + \hat{\sigma}^2_0 (1 - exp(-\hat{\theta}h))\]
with \emph{practical range} 
\[h^* = \frac{log(\hat{\sigma}^2_0/0.05)} {\hat{\theta}}   \]
where $\hat{\theta}$ is a parameter to be fitted.

\subsection*{Prediction of  spatially structured variables at unmeasured points}
There are different methods to predict a value at locations within a continuous area (random field) where no measurement is available. With \emph{kriging} the value of one or more characteristic(s) $Z$ at a given point $\mathbf{s}_0$ is predicted from the known values of this characteristic at nearby points. The result is a weighted average of all known points in the area (global kriging) or of a selection of neighbouring points (local kriging), that is a linear combination of known data. Here the weighing is provided by the exponential semi-variogram model and depends on the spatial correlation structure, in particular on the distance between points such that points at distances above the range have weight 0.
\subsubsection*{Ordinary kriging}
Under the assumption of a constant but unknown mean $\mu$ of the two-dimensional random field, we perform \emph{ordinary kriging} to predict unobserved values $Z(\bm{s}_0)$ at (usually unobserved) locations $\mathbf{s}_0 = [x_0, y_0]$. The assumed model is: 

\[\mathbf{Z(s)} = \mu \bm{1} + \bm{e(s)}, \quad \bm{e(s)} \sim (\bm{0,\Sigma}),  \] 
with $\mathbf{Z(s)}$ representing the observed spatial data $\left[Z(\bm{s}_1), Z(\bm{s}_2), \dots , Z(\bm{s}_n) \right] $, and
\begin{align*}
E(\mathbf{Z(s)}) &= \mu \bm{1}\\    
Var(\mathbf{Z(s)}) &= \bm{\Sigma} 
\end{align*}

\noindent
Assumptions:
\begin{itemize}
\item the $\mu$ is unknown but constant over the random field
\item $\bm{\Sigma}$ is assumed to be known in this model. In practical applications this is often not the case, and it has to be estimated based on the semi-variogram model. 
\item $Var(Z(\bm{s}_0)) = \sigma^2 = C(\bm{0})$
\item $Z(\mathbf{s})$ is a secondary stationary process (for the benefit of semi-variogram estimation).
\end{itemize}
The ordinary kriging predictor at a location $\mathbf{s_0}$ is 
\[p_{ok}(\bm{Z;s}_0) = \bm{\lambda'Z(s)},\]
with $\bm{Z}$ representing the observed data and 
$\bm{\lambda}$ the matrix of kriging weights. And these weights are determined by 
\begin{enumerate}
    \item The vector $\bm{\sigma} = Cov \left[Z(\bm{s}_0), \bm{Z(s)}   \right]$
    \item The variance-covariance matrix $\bm{\Sigma} = Var\left[ \bm{Z(s)} \right]$ or the corresponding inverse.
\end{enumerate}

\subsubsection*{Co-kriging}
The kriging model can be extended to the simultaneous prediction of  several variables which are  spatially correlated with each other \cite{Schabenberger.2005}.  This is typically the case when constructing an index based on a number of related neighbourhood characteristics.

In this case, in addition to a semi-variogram for every variable, a cross-semi-variogram for each pair of variables can be modelled. The fact that cross-covariance functions have to be specified in such a way that for every selection of locations a positive definite covariance matrix results presents a challenge. We use the linear model of co-regionalisation (LMC) implemented in the {\it gstat} package in R \cite{Goulard.1992, R, Graler.2020, Pebesma.2004}.

The different variables can be measured at identical locations (for example, when they are from the same survey) which is called \emph{collocated} or \emph{isotopic} measures \cite{Wackernagel.2002} or each at different locations (when from different data sources which may use different spatial scales), which is called \emph{heterotopic}. If all variables are indicators of the same spatial process, spatial correlations can be assumed in both cases.

A question that the simulation study needs to answer is if in order to predict a linear combination of variables (index) it is better to take the correlation structure between variables into account by using a co-kriging model with a more complex estimation procedure or to predict the variables separately. To compute the index value, we combine the predicted values at a specified location as unweighed sum or mean of the three variables in both cases.

\subsection*{Simulation study}

The simulation scenarios are based on previous studies which provide us with  ranges for the semi-variogram and a number of neighbours within this range which corresponds to the kind of data which might be expected if some form of local sampling is performed (e.g. spatial random walk). The size of the sample surface and the density  of the simulated data points are based on the overall sample size available to estimate an exponential  semi-variogram model and the number of points available at a given range (radius) to predict values from the data available.

We performed two families of simulations. One in which the data is used to estimate the semi-variogram model for kriging and one in which the range is given a priori based on the theoretical considerations. The limitation of performing a simulation in which the semi-variogram model is estimated, is that there is no possibility to check that the semi-variogram model does not fit the empirical semi-variogram. When this occurs, the person fitting the model would change the starting parameter for the estimation or the maximal distance defining the empirical semi-variogram. This cannot be done in a simulation and leads to artificially increased variability. Therefore we include a check of validity in which only models with "reasonable" shape parameter are considered. For the simulation with fixed range we use ranges which may differ from the true range of the simulation process. The simulation parameters are provided below and in Table \ref{tab:tableunivar}. 

The aim of the simulation study is twofold: 1. evaluate the reliability of prediction using a semi-variogram kriging model when the semi-variogram must be estimated from the data and only a limited number of observations is available at small distances and 2. find out whether using a co-variogram to predict  several  spatially  correlated measures improves the prediction of a summary index compared to predicting the measures separately. The simulation model reproduce the type of data analysed in other studies (e.g. \cite{Breckenkamp.2021} but also based on theoretical considerations in which only the local spatial scale is relevant and spatial processes at wider ranges are ignored.

\subsubsection*{Simulation of true values}
We start by simulating  random fields. We assume a second order stationary process with an exponential semi-variogram function in all simulated fields with varied ranges. The variance of the considered variable was set to 1. Overall, twelve random fields were simulated on regular grids with different dimensionalities and ranges.  From these realisations we sampled the observed values used to estimate the parameters of the semi-variogram model and the kriging predictions. 
Different sampling densities were simulated by varying the dimensionality of the simulated fields to be 8,000m, 10,000m, and 15,000m and for each field size sampling 3 different numbers of points to result in three comparable densities. Half of the random fields were simulated without a nugget effect and the other half with a nugget effect of 0.2 (20\% of the true variance).
For the co-regionalised (multivariate) model, three variables were simulated with varying pairwise correlations between these variables. The data were simulated with the R package \emph{RandomFields} \cite {Schlather.2015}. 

\subsubsection*{Sampling from the random fields}
Given the sample size $n$ corresponding to different densities and the ranges we repeatedly (5,000 times) sampled sets of $n$ points from the corresponding simulated random fields. These points were used to estimate the parameters of the exponential semi-variogram model first and the models were then used to obtain kriging estimates at the selected 200 test points. In a second variant, kriging was based on semi-variograms with parameters fixed at the true values for range and nugget. 
For co-kriging we distinguished two cases: First, we sampled collocated variables where all three variables were measured at the same locations with varying the densities (and pairwise correlations). Second, the three variables were sampled from different locations, each with a different density (again from random fields with varying pairwise correlations). 

\subsubsection*{Selecting fixed test points for prediction}
From the simulated random fields, we randomly chose 200 fixed test points for each extent of random field on which to obtain kriging estimates. The same 200 test points were chosen for different parameter values for each simulated field. 
Ordinary kriging (function \texttt{krige} from \emph{gstat}) was used to predict the values at these 200 fixed test points based on the  values at the sampled locations for each sample and on the estimated or fixed exponential semi-variogram model from this sample, with a maximal distance for the estimation of the empirical semi-variogram equal to 1,000 (about 1.5 times to 3 times the true range).

\subsubsection*{Varying parameters}
To compare the quality of predictions between different scenarios we chose the parameters:  
\begin{itemize}
\item \texttt{dimensionality of random fields}: Simulated quadratic random fields with the dimensionalities of 8,000m, 10,000m, and 15,000m were used to sample from.
\item \texttt{ranges of semi-variogram}: Ranges were 300 and 600m. 
\item \texttt{nugget effects of semi-variogram}:  The nugget was set to be 0 (i.e. no nugget effect) or 0.2.
\item \texttt{number of sampled points/density}: To arrive at three comparable expected densities of observed points within a specific radius, we then chose the number of sampled points dependent on the dimensionality of the random field we sampled from. Numbers of points were: 650, 1,300, 2,300 for the smallest fields ($8,000m^2$), 1,000, 2,000 and 3,500 for the medium ($10,000m^2$), and 2,500, 5,000, and 8,000 points for the largest fields ($15,000m^2$). This resulted in expected numbers of points within a distance of 250m of about 2, 4, and 7 for each size of random field. 
{\it Increasing the range for a fixed distance between points means that the number of pairs available to estimate the spatial correlation structure increases. Therefore, we expect that the kriging model based on the parametric semi-variogram will better perform for larger ranges.}   
\item \texttt{semi-variogram parameters}: Parameters of exponential semi-variogram models used for kriging were either estimated from the sampled points or fixed at the true values.  
\end{itemize}

\noindent
For the multivariate co-kriging models we also varied
\begin{itemize}
\item  the pairwise correlations between the three different variables to be $r= 0.1$, $r=0.5$, or $r=0.9$  
\item and the type of predictions being univariate ordinary kriging (3 variables predicted separately) versus co-kriging (3 variables predicted simultaneously based on co-variograms). 
\end{itemize}

\noindent
Twelve sets of results were generated for each size of random field with estimated and another 12 with fixed semi-variograms. Overall, 72 generated result sets for different parameter combinations (dimensionality, range, density, nugget, semi-variogram parameters) each consisting of repeated kriging predictions for 200 testpoint locations were included in our analyses in the univariate (single measure) case. For the multivariate case only fixed semi-variograms were used, kriging was repeated 1,000 times, the range was set to 600, and the maximum number of nearby points to include for kriging was fixed to be 50 within a maximum radius of 1,000m. When the three variables were sampled from different locations density was varied between the variables. 
	
\subsection*{Evaluation}
The precision of the estimates at a given point is evaluated using the empirical standard error defined as the standard deviation of all the estimated values at this point over all 5,000  samples from a given simulation scenario. 
The 200 test points are classified within the quintile categories of the outcome distribution. Then the reliability of the estimation method is based on the proportion of estimated points which fall into the original quintile as well as the percentage which fall in a neighbouring quintile.
The evaluation of the multivariate approach mainly focuses on the benefit of using a co-variogram model as opposed to estimate the different variables independently of each other.

\subsection*{Case study}
\begin{figure}[h!]
 \centering
\includegraphics[height = 0.4\textheight]{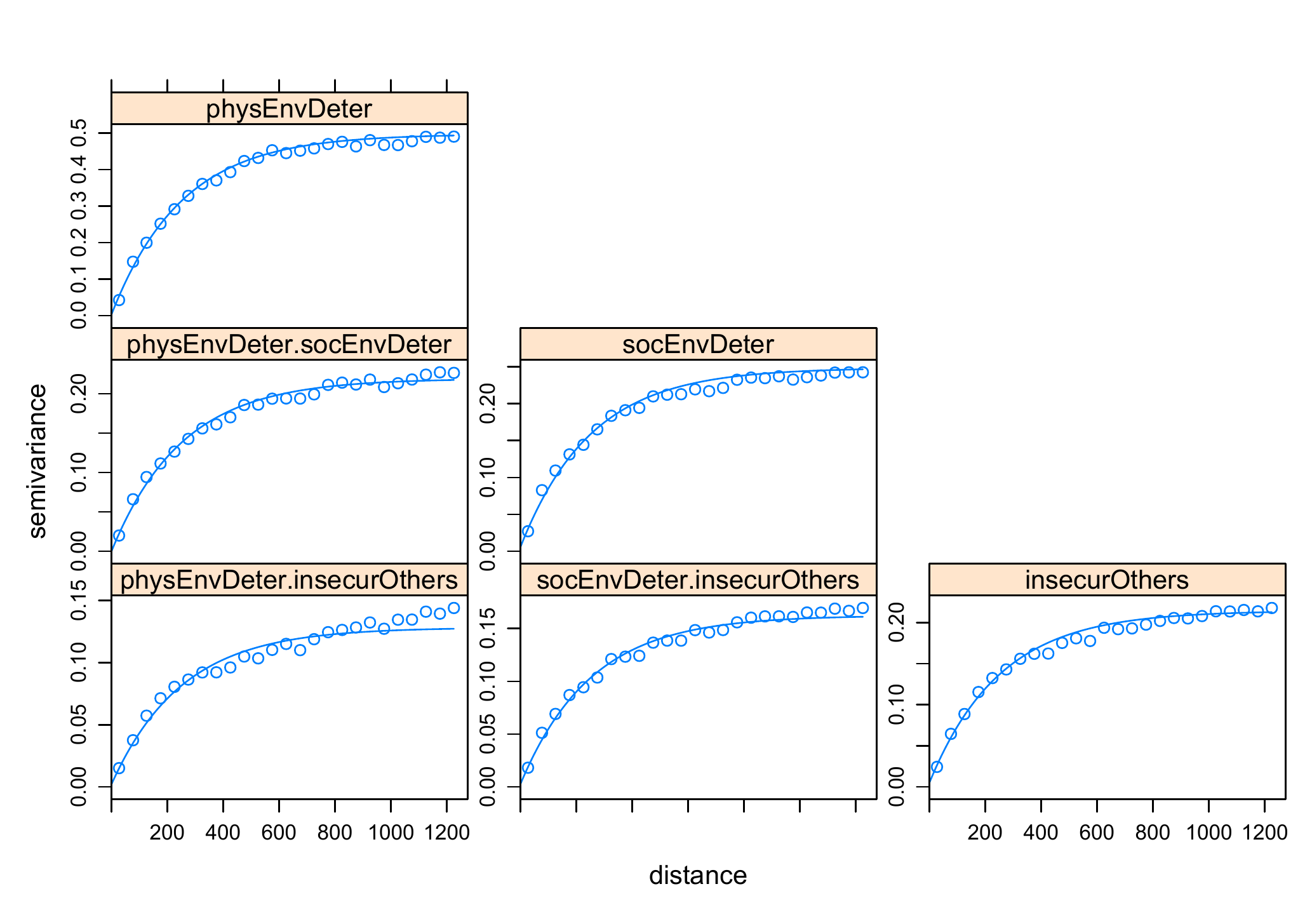}
\caption{Co-variograms estimated from all available RECORD data: observed semivariances and fitted exponential models (semi-variograms on the diagonal and cross-variograms off diagonal)}   
\label{fig:covario}
\end{figure}

We used data from the “Residential Environment and CORonary heart Disease”  study (RECORD)\cite{Chaix.2012} to illustrate the method described above. This cohort study conducted in the Paris Ile-de-France region in 2007 and 2008 with 7,290 participants contains in particular data on a range of perceived neighborhood characteristics as well as geo-coded addresses of participants. 
We selected three variables from the dataset defined as scores based on questionnaire items on perceived neighborhood characteristics. These three scores have a spatial correlation structure  and are also spatially correlated with each-other (see Figure \ref{fig:covario} for the co-variograms). 
The three scores on perceived physical and social deterioration as well as  insecurity from others in the neighborhood were selected based on high inter-correlations in the study which makes them candidates for measures of a homogeneous index and likely to be generated by the same spatial process (an assumption of linear models of co-regionalisation). 
We randomly chose 200 participants with known residential addresses for which the perceived deterioration and insecurity of their neighbourhood was available. We predicted these values using several semi-variogram models (see below), and we compare the predicted with the true values. 

First, we used all available data to fit exponential semi-variogram models for the three variables and their co-variograms by employing the linear model of co-regionalization. A good fit for the three exponential models was obtained for maximal distances of the empirical semi-variogram of 1,000m to 1,250m (see Figure \ref{fig:covario}). Deviations from the shape of the model at larger distances indicated the existence of additional spatial processes on larger scales. We therefore only used points within a distance up to 1,250m to fit the semi-variogram model parameters, resulting in an estimated range of 756m (scale parameter of 252 for exponential model) common to the three variables.  In a second stage, we sampled varied numbers of observations that we used both to estimate the semi-variogram exponential parameters and to predict values at the 200 points.

The kriging predictors were obtained using a number of randomly sampled known points varying between 500 and 7,090 (total number of observations minus the 200 points to be predicted). The minimum number of used neighbouring points was fixed to be 50.  

The mean of the kriging predictions of the three neighborhood variables at a given location was considered as representing the index value at this point. We then compare the estimated values to the true observed values at the 200 test points. The proportions of correct or neighbouring index quintiles as well as deviations from the true index values are reported for the varying parameter combinations.

\section*{Results}
\subsection*{Simulation: Reliability of single point estimation}
We predicited the values at 200 fixed points (test points) by ordinary kriging based on 5,000 different samples (1,000 in the multivariate case) for each combination of parameters.  
The values for the different reliability measures are summarized per point and per parameter combination.    

\begin{figure}[h!] 
 \centering
\includegraphics[height = 0.4\textheight]{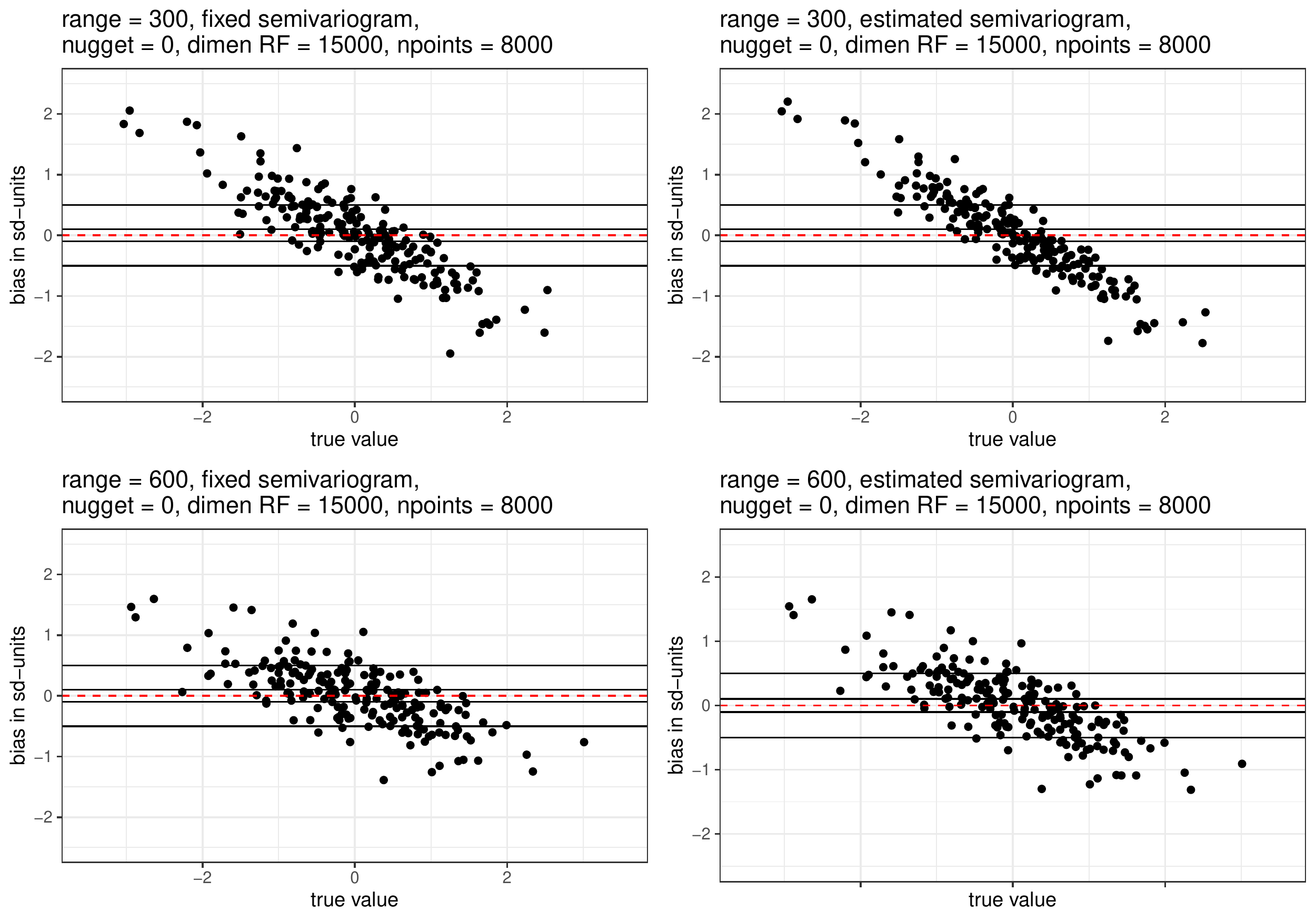}
\caption{Distribution of bias at 200 testpoint locations by different semi-variogram ranges and fixed versus estimated semi-variograms, mean bias per point expressed in standard deviation units} 
\label{fig:bias.range}
 \end{figure}

In Figure \ref{fig:bias.range} the plots display the means summarized over the 5,000 samples per test-point for the 200 locations.
That figure represents the true values at the 200 estimated points against the bias of the estimated kriging values (mean difference between kriging estimate and true value of the simulated random field) depending on the true range of the semi-variogram model and based on fixed versus estimated semi-variograms. The bias clearly depends on the magnitude of the true value but this dependency decreases with increasing range. For smaller ranges a reduction to the mean for extreme true values is apparent. With increasing range, the bias overall decreases (see also Table \ref{tab:tableunivar}).

\begin{figure}[h!]
 \centering
\includegraphics[height = 0.4\textheight]{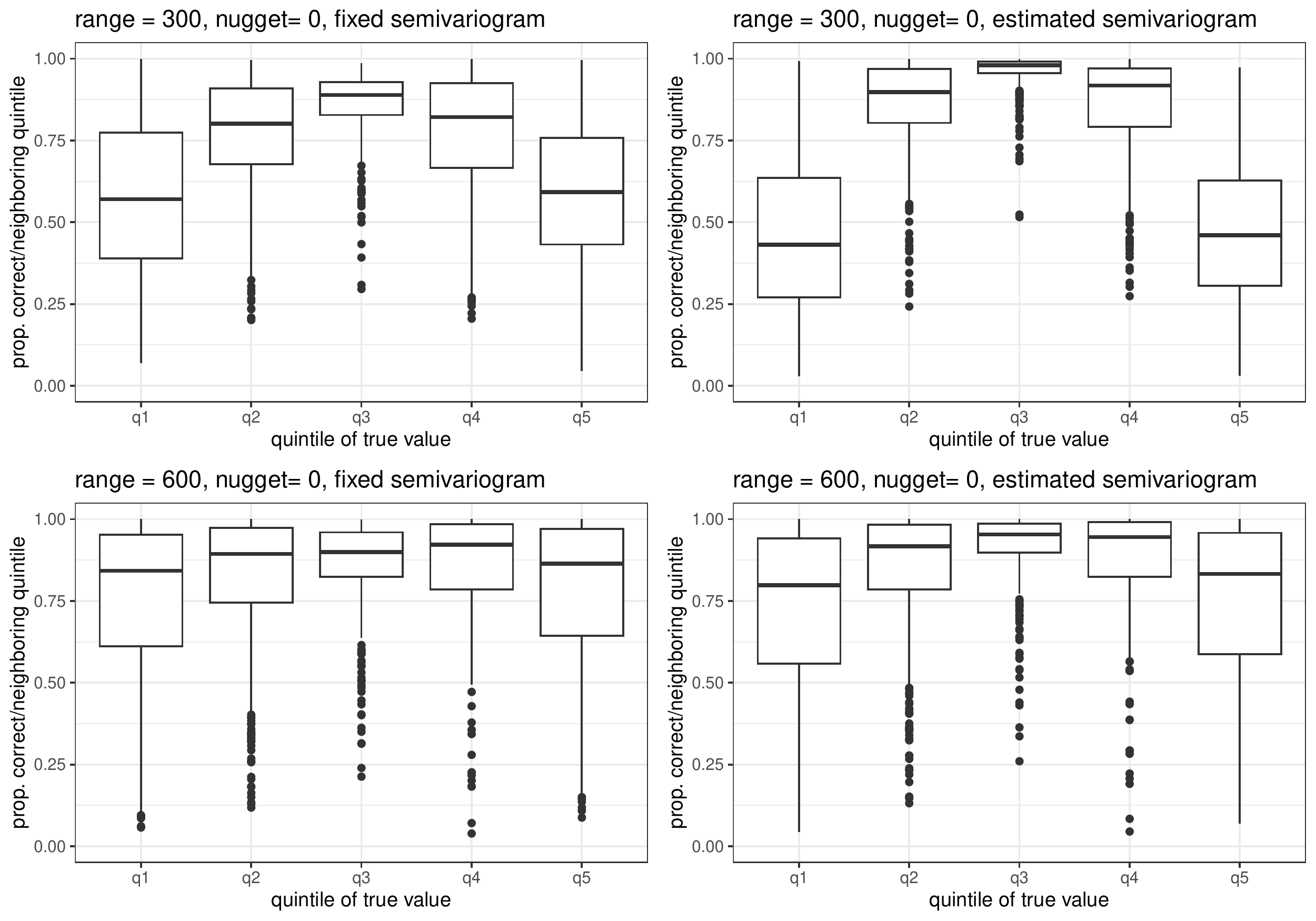}
\caption{Proportion of predicted points in the correct or a neighbouring quintile grouped by true quintile value}  
\label{fig:quintiles}
\end{figure}

Figure \ref{fig:quintiles} shows for each quintile the proportion of predicted values which are in the correct quintile, or a neighbouring quantile. With increasing range the proportion of estimations falling in the correct or neighbouring quintile increases, first and foremost for the lowest and highest quintile. 
Aside from that, a shift of predictions to the mean was more marked when the semi-variogram parameters were estimated from the sampled observed points than when fixed to the true values, and this being especially pronounced at a smaller range.
Results for simulations with a nugget effect of 0.2 look very similar (see Table \ref{tab:tableunivar}).

In Table \ref{tab:tableunivar} the values are further summarised over the 200 test-points, showing mean, median, and SD of the distributions.  The different quality measures are given for increasing extent of the simulated random field, three different sample sizes (resulting in 3 densities comparable between random fields of different dimensionalities), 2 ranges, with and without nugget effect, and  semi-variogram parameters either fixed or estimated. 
A larger range and an increasing density were apparently related to smaller prediction errors (MSE) and bias, and, accordingly, with higher poroportions of correct predictions. 
Having a larger range means than the information held in the observed point can be used at a larger distance. Increasing the sample size and, therefore, the density has two advantages, more points are available for the estimation of the model leading to a better semi-variogram model, and there are more nearby points to be used for the predictions. 

A higher dimensionality came along with a reduction in bias but no clear advantages in terms of other measures. 
Introducing a nugget effect led to less accurate predictions of quintiles and a markedly increased MSE. 
These patterns of results were very similar for fixed and empirically estimated semi-variograms. Only the precision increased with higher density or range with the fixed semi-variogram, and proportion of the correctly predicted quintiles was slightly better in this case. Overall, the comparison of both variants shows better performance with estimated semi-variogram parameters, especially considering the prediction errors and precision. 
On average, up to 80\% of all predictions were found in the correct or its neighbouring quintile, but the exact correct quintile could only be predicted in 36\% of cases in maximum.  


\subsection*{Results of multivariate simulation and prediction}
Values for the three variables were predicted at the same 200 test point locations as in the univariate case. At each test point an index value was computed as sum of the three predicted variables and this was compared to the sum of the true simulated values. All results relate to these index values. 
Besides the variable correlations which were varied in the simulations, the multivariate kriging scenarios differed from the univariate ones in fixing the maximum number of neighbourhood points used for local kriging to 50 and the number of repeated sampling and prediction to 1,000. Both was done to account for the increased complexity of the model and related time as well as memory capacity requirements.    
Results are displayed in tables \ref{tab:multi1} and \ref{tab:multi2}).   

\paragraph*{Comparison univariate versus multivariate.}
The patterns of results were virtually similar for the separate univariate predictions and co-kriging predictions, when all 3 variables were measured at the same locations (collocated).  
Some differences were seen in the multi-located (heterotopic) case. Here, the bias of predictions was smaller for less correlated variables in the univariate case but slightly smaller with a high correlation for the co-kriging results. The proportion of exactly correct predictions of quintiles increased by 5 percentage points in the univariate but 11 percentage points in the co-kriging predictions with multi-located variables. However, this pattern was not evident when considering the proportions including neighbouring quintiles. 

\paragraph*{Comparison point collocation versus multi-location.}
We compare the results from Tables \ref{tab:multi1} and \ref{tab:multi2}. Scenarios with three variables observed at different locations show better results than for variables observed at the same locations. We see up to 82\% of points in the correct or neighborhood quintile for multi-located observations but only up to 72\% of collocated observations for the univariate method and 83\% vs. 72\% for the multivariate method. Bias and MSE were also clearly greater in the collocated than in the multi-located case.

\paragraph*{Effect of the correlation between variables.}
The more the variables are correlated the more information can be gained from the others and, therefore, improvements in the predictions in the case that observations are at different locations for different variables were to be expected. However, different correlated variables from the same location share much information, increasing with size of the correlation. Therefore, the information gain is decreasing with increasing correlations. This is confirmed by the simulations which show that increasing the correlation between the variables from 0.1 to 0.9 led to a decrease in the proportions of prediction in the right or neighbouring quintile (71 to 65\%) in the collocated case, both for univariate and multivariate predictions. 
When the points are multi-located only the percentage of exactly predicted quintiles increased with increasing correlation (35 to 40\% univariate and 36 to 47\% multivariate). This seems less relevant, since overall the proportion of exact predictions still seems too low for practical applications.  

\paragraph*{Further results}
Introducing a nugget effect lowered prediction quality in all scenarios in terms of correctly predicted qunitiles and prediction error. Bias was mainly unaffected in the collocated case but was decreased in models without nugget effect for multi-located observations. 

Density, which also corresponds to the number of sampled observations, did not affect prediction quality. 

In the collocated case, the extent of the random field made no clear difference in the proportion of predicted quintiles. Bias was minimal for the smallest field. With multi-located obervations we observed the best quintile prediction as well as smallest prediction errors with the medium sized field and comparable smaller biases either in the medium or small field. 

\subsection*{Case study}
The results of the case study based on RECORD data are shown in Tables \ref{tab:record1} and \ref{tab:record2}. The corresponding semi-variogram parameters estimated from different numbers of sampled points and varying the maximal distance for semi-variograms are provided in Table \ref{tab.semi.var.Record}. The three variables considered are approximately normally distributed.

The mean of true index values over the 200 test points is $-0.094$, $median =  -0.091$, and $SD = 0.624$. The corresponding kriging predicted values were $mean = -0.097$, $median = -0.102$, and $SD = 0.478$ for the univariate version when using all available points for the estimation of the semi-variograms. The values were $mean = -0.098$, $median = -0.107$, and $SD = 0.406$ with a co-variogram based on all available points. For results based on varying (co-)semi-variograms see Tables \ref{tab:record2} and \ref{tab.semi.var.Record}. 

\paragraph{Semi-variogram estimated using all available observations.}
In Table \ref{tab:record1} the measures of the quality of estimates are presented with varying the number of sampled known points used for kriging but using the semi-variogram obtained from all 7,290 available points. Here we see that multivariate and univariate approaches are equivalent in terms of quality and only large sample sizes provide a high percentage of prediction in the correct or neighbouring quintile: 88\% for 500 points to 95\% for 7,090 known points. The bias ranging from 20\% of the index mean to 5\% with half of predictions having less than 1\% bias for 7,090 known points.

\paragraph{Semi-variograms estimated using the subset of observations available for prediction.}
In table \ref{tab:record2} the measures of the quality of predicted values from semi-variograms based on only the number of sampled points are presented and the corresponding semi-variogram models are given in Table \ref{tab.semi.var.Record}. Here, the increase in the quality of prediction with the number of sampled known points is more pronounced in the multivariate case. That is, specifically the realistic estimation of the cross-variograms seems to depend on the number of known points. However, the difference between univariate and multivariate kriging is mainly seen up to 2,000 known points.

\section*{Discussion}
We have shown the usability and reliability of kriging methods to create geo-located indices of neighbourhood characteristics based on survey data. The particularity of the method compared to the existing literature is that it does not rely on predefined geographical units, thus allowing for very fine spatial granulation of characteristics. The focus of our study is  on geo-located data sources (survey data) for which only sparse data at small distance to the points being predicted is available, thus offering limited information for prediction. The benefit of such data however for the study of small-area health inequalities is to provide data on very local neighbourhood characteristics (which may be associated with health outcomes) relative to individual health data otherwise rarely available, while geo-coded data on health outcome may be increasingly available.

We based our kriging models on semi-variograms commonly seen in social-epidemiological studies of neighbourhood effects on health in other context \cite{Breckenkamp.2021, Sauzet.2021}. Here the prediction at one location is based on the spatial correlation structure between observed values around the non-observed location. The simulation scenarios were chosen to reflect the correlation structure of existing data as well as a sparse number of observations at small distance to the location at which predictions are performed (from about 10 to 40 observations). The mean bias ranges among the simulated scenarios from 10 to 3\% of the standard deviation of the outcome providing on average a reliable prediction of the true value.  
We also looked at data categorised in quintiles with a median over the simulations having 80\% or more predictions in the right quintile or its neighbour. Therefore the method may provide a useful indicator of neighbourhood characteristic at one location with the limitation that the variability may be reduced if too much reduction to the mean occurs. Nevertheless, this is a better than predicting the same values for a whole area.  

We presented a case study using a large study with about 7,000 geo-coded observations from which we drew samples of various sizes.  With using a sample of only 500 data points available for prediction (average of 3.2 (SD 2.6) points within the range of predicted locations), 86\% of predictions were in the correct or neighbouring quantile going up to 95\% when using all the data available (average of 45.1 (SD 33.8) points within the range of predicted locations). Showing that the method provided satisfactory predictions with very little neighbouring points.

As expected,  larger sample sizes and more data at small distances will both provide  better estimates of the semi-variogram model parameters. But relevant for the predictions  is that there are enough points within the range radius and a small nugget effect. A nugget effect means that only a part of the information available in the points within the range radius will be used for prediction.  The role of the range of the semi-variogram model is double. If the true range of the exponential model is larger (see simulation results) then more observed values will be (correctly) used for prediction. However, if the estimated range is larger than the true range (as seen in some sub-samples of the case study)  then some observed values will be used for prediction which are actually uncorrelated with predicted points, thus reducing the quality of predictions. 

We performed prediction based on estimated semi-variograms as well as with a semi-variogram given a priori. On average estimation provided better results, probably because  the high variability of the simulated data meant that the "true" semi-variogram did not always fit the specific sample data well. However, there may be situations where it makes sense to use a range and a nugget effect based on theory or previous work. In that case the weight put on each point is not based on the correlation structure observed in the data. 

Another aspect of our investigations was whether using co-variogram models, which take the spatial correlation structure between different variables included in the index into account, would provide a better prediction of indices (as linear combinations of predicted values) than the linear combination of separately predicted values. The advantage of using the information gained by including the spatial correlation between different variables has to be balanced with the possibility of problems with the estimation of the co-variogram model. In our case study, if all points available were used to estimate the models (thus providing a rather good semi-variogram and probably co-variogram) then the multivariate and the univariate approaches lead to similar results. Whereas, as in the simulations, if the semi-variogram models are estimated using a smaller sample of points then the  univariate approach provides better results. Therefore in general it does not seem to be hugely beneficial to use co-variogram models and with increasing number of variables the difficulties to obtain estimates increases. If the different variables are not observed at the same points, e.g. if the data come form different sources, then a co-variogram will provide better estimates. 

Given the usefulness of spatial methods based on semi-variograms to provide neighbourhood indicators not relying on administrative areas presented here or elsewhere, it is important that designers of social surveys take geographical proximity into account in their sampling frame. This will allow to obtain estimates of neighbourhood characteristics which are not based on the perception of one person only but on the perception of several neighbours. Our case study was based on a an epidemiological study with a large sample size with participants selected in randomly chosen "arrondissements" or communes of the Paris area. As a consequence  the number of  neighbours within 750m ranges from 10 to 120, thus providing very good estimates at non observed locations. A major drawback is the presence of observation deserts.  Having enough data for kriging provides another possible application of the method proposed. We can obtain a measure of neighbourhood perception of an individual and an "objective" measure of neighbourhood characteristic with the kriging predicted values provided by the neighbours.

A limitation of this work is the lack of estimation of prediction error. The estimation of semi-variogram parameters standard errors is difficult, in particular because of the type of data available (an empirical semi-variogram) and that the estimates are skewed (see \cite{Dyck.2022} for more details). Further work is needed to provide a reliable measure of precision. This is particularly important if an indicator is based on a large number of variables bringing each a source of uncertainty. A direction of investigation is the work of Lopiano et al on misaligned data \cite{Lopiano.2011} used in regression models. 

\subsection*{Conclusion}
Provided that survey data and study data close enough to each other in sufficient numbers are available, survey data with geo-coded spatially correlated neighborhood characteristics can be used to predict values for these variables at nearby locations using semi-variogram kriging models. This enables, for example, the creation of bespoke indices for the study of small-area determinants of health which are valid at very small spatial scales and comprising relevant dimensions for the population under study.  

\section*{Acknowledgement}
We thanks the two reviewers for their very valuable comments, which helped us improve this work considerably.
We wish to thank Basile Chaix for providing us with the data of the RECORD study for the case study.
This research was funded by the German Research Foundation, DFG (Research unit PH-LENS, FOR 2928 - 409654512, Sub-project DEPRIV, PI: Oliver Razum and Odile Sauzet).   




\newpage
\section*{Tables}

 \begin{table}[htbp]
   \centering
   \caption{Expected number of points within radius (mean and SD of thev no. of points over the 200 test points per sample):
}
     \begin{tabular}{rrrrrrr}
     \toprule
     \multicolumn{1}{l}{Radius:} & \multicolumn{1}{l}{250m} & \multicolumn{1}{l}{300m} & \multicolumn{1}{l}{600m} & \multicolumn{1}{l}{750m} & \multicolumn{1}{l}{1000m} & \multicolumn{1}{l}{1250m} \\
     \midrule
     \multicolumn{1}{l}{\textbf{No. of sampled points:}} &     &     &     &     &     &  \\
     \multicolumn{1}{l}{\textit{random field 8,000 $\text{m}^2$:}} &     &     &     &     &     &  \\
     650 & 1.9 (1.3) & 2.7 (1.6) & 10.6 (3.5) & 16.2 (4.8) & 28.1 (7.5) & 42.5 (10.6) \\
     1300 & 4.0 (2.0) & 5.8 (2.5) & 21.8 (5.6) & 33.3 (7.4) & 56.5 (11.7) & 85.4 (18.0) \\
     2300 & 6.8 (2.6) & 9.7 (3.0) & 37.4 (8.0) & 57.7 (11.7) & 99.8 (19.5) & 150.4 (30.7) \\
     \multicolumn{1}{l}{\textit{random field 10,000 $\text{m}^2$:}} &     &     &     &     &     &  \\
     1000 & 2.1 (1.4) & 2.9 (1.7) & 11.0 (3.5) & 16.7 (4.8) & 28.8 (7.4) & 44.0 (10.6) \\
     2000 & 4.1 (1.9) & 5.8 (2.3) & 21.1 (5.4) & 32.7 (7.8) & 56.8 (14.3) & 86.8 (22.2) \\
     3500 & 7.0 (2.8) & 9.9 (3.6) & 37.6 (7.5) & 57.6 (10.6) & 99.8 (18.5) & 151.6 (30.5) \\
     \multicolumn{1}{l}{\textit{random field 15,000 $\text{m}^2$:}} &     &     &     &     &     &  \\
     2500 & 2.0 (1.4) & 3.0 (1.6) & 12.2 (3.5) & 18.7 (4.4) & 32.9 (6.8) & 50.6 (9.6) \\
     5000 & 4.1 (2.1) & 6.1 (2.5) & 23.9 (5.4) & 37.5 (7.4) & 66.0 (12.4) & 101.7 (18.8) \\
     8000 & 6.9 (2.5) & 9.8 (3.1) & 38.3 (6.3) & 59.7 (9.8) & 104.6 (16.3) & 160.7 (25.3) \\
     \bottomrule
     \end{tabular}%
     \label{tab:npoints}%
 \end{table}%

\begin{table}[htbp]
   \centering
   \caption{Quality of prediction in different kriging scenarios (univariate  simulations)} 
     \begin{tabular}{rrrrrrrrrrrrrrr}
     \toprule
         & \multicolumn{3}{c}{poportion corr. Q.} &     & \multicolumn{3}{c}{corr./neighbor Q.} &     & \multicolumn{1}{c}{mean} &     & \multicolumn{1}{c}{mean} &     & \multicolumn{1}{c}{mean} &  \\
\cmidrule{2-4}\cmidrule{6-8}         & \multicolumn{1}{c}{mean} & \multicolumn{1}{c}{SD} & \multicolumn{1}{c}{med} &     & \multicolumn{1}{c}{mean} & \multicolumn{1}{c}{SD} & \multicolumn{1}{c}{med} &     & \multicolumn{1}{c}{Bias} &     & \multicolumn{1}{c}{SE} &     & \multicolumn{1}{c}{MSE} &  \\
     \multicolumn{14}{l}{\textit{\textbf{Semi-variogram parameters estimated from sampled (observed) points}}} &  \\
     \midrule
     \multicolumn{15}{l}{\textit{Extent of Random Field (dimensionality in $m^2$)}} \\
     \midrule
     8000 & 0.33 & 0.23 & 0.30 &     & 0.78 & 0.26 & 0.90 &     & 0.081 &     & 0.357 &     & 0.752 &  \\
     10000 & 0.29 & 0.22 & 0.24 &     & 0.73 & 0.28 & 0.85 &     & 0.042 &     & 0.362 &     & 0.854 &  \\
     15000 & 0.32 & 0.22 & 0.31 &     & 0.78 & 0.25 & 0.89 &     & 0.007 &     & 0.352 &     & 0.712 &  \\
     \midrule
     \multicolumn{15}{l}{\textit{Range}} \\
     \midrule
     300 & 0.27 & 0.23 & 0.21 &     & 0.72 & 0.29 & 0.85 &     & 0.048 &     & 0.343 &     & 0.885 &  \\
     600 & 0.35 & 0.22 & 0.36 &     & 0.80 & 0.23 & 0.90 &     & 0.038 &     & 0.371 &     & 0.660 &  \\
     \midrule
     \multicolumn{15}{l}{\textit{Density (no. of known points sampled)}} \\
     \midrule
     2   & 0.28 & 0.23 & 0.21 &     & 0.72 & 0.29 & 0.86 &     & 0.045 &     & 0.342 &     & 0.870 &  \\
     4   & 0.31 & 0.22 & 0.29 &     & 0.77 & 0.26 & 0.88 &     & 0.044 &     & 0.363 &     & 0.767 &  \\
     7   & 0.35 & 0.23 & 0.34 &     & 0.80 & 0.24 & 0.90 &     & 0.041 &     & 0.366 &     & 0.681 &  \\
     \midrule
     \multicolumn{15}{l}{\textit{Nugget}} \\
     \midrule
     0   & 0.33 & 0.22 & 0.32 &     & 0.78 & 0.25 & 0.89 &     & 0.042 &     & 0.356 &     & 0.653 &  \\
     0.2 & 0.29 & 0.23 & 0.25 &     & 0.75 & 0.28 & 0.87 &     & 0.044 &     & 0.358 &     & 0.892 &  \\
     \midrule
     \multicolumn{15}{l}{\textit{\textbf{Semi-variogram parameters fixed}}} \\
     \midrule
     \multicolumn{15}{l}{\textit{Extent of Random Field (dimensionality in $m^2$)}} \\
     \midrule
     8000 & 0.34 & 0.20 & 0.35 &     & 0.77 & 0.23 & 0.86 &     & 0.084 &     & 0.436 &     & 0.781 &  \\
     10000 & 0.29 & 0.20 & 0.28 &     & 0.71 & 0.25 & 0.79 &     & 0.049 &     & 0.446 &     & 0.924 &  \\
     15000 & 0.34 & 0.20 & 0.34 &     & 0.77 & 0.23 & 0.86 &     & 0.006 &     & 0.430 &     & 0.747 &  \\
     \midrule
     \multicolumn{15}{l}{\textit{Range}} \\
     \midrule
     300 & 0.28 & 0.16 & 0.29 &     & 0.71 & 0.24 & 0.78 &     & 0.052 &     & 0.473 &     & 0.955 &  \\
     600 & 0.36 & 0.23 & 0.36 &     & 0.80 & 0.23 & 0.89 &     & 0.041 &     & 0.402 &     & 0.680 &  \\
     \midrule
     \multicolumn{15}{l}{\textit{Density (no. of known points sampled)}} \\
     \midrule
     2   & 0.29 & 0.17 & 0.29 &     & 0.71 & 0.24 & 0.78 &     & 0.051 &     & 0.470 &     & 0.938 &  \\
     4   & 0.32 & 0.20 & 0.33 &     & 0.76 & 0.24 & 0.84 &     & 0.046 &     & 0.438 &     & 0.809 &  \\
     7   & 0.35 & 0.23 & 0.36 &     & 0.79 & 0.23 & 0.89 &     & 0.042 &     & 0.404 &     & 0.706 &  \\
     \midrule
     \multicolumn{15}{l}{\textit{Nugget}} \\
     \midrule
     0   & 0.34 & 0.20 & 0.34 &     & 0.77 & 0.22 & 0.84 &     & 0.045 &     & 0.454 &     & 0.700 &  \\
     0.2 & 0.30 & 0.20 & 0.29 &     & 0.74 & 0.25 & 0.83 &     & 0.048 &     & 0.420 &     & 0.935 &  \\
     \bottomrule
     \end{tabular}%
   \caption*{\footnotesize{Note: Values are summarized over the 200 testpoints; SD: standard deviation; med: median; prop.corr. Q.: proportion of predicted values in the correct quintile; corr./neighbor. Q.; proportion of predicted values in the correct or a neighboring quintile; mean bias: mean of predicted minus true value in SD units of true outcome; SE: mean standard error of predicted kriging values; MSE: mean squared error of kriged values; No. of points: number of sampled points used for kriging; Range: range of semi-variogram models used for simulation of random fields; Density: density of sampled observed points (based on different numbers of points sampled) }}
   \label{tab:tableunivar}
 \end{table}%

 \begin{table}[h!tbp]
   \centering
    \caption{ Quality of prediction in different kriging scenarios: multivariate simulation with 3 variables observed at different locations (1,000 known points)}  
      \begin{tabular}{rrrrrrrrrrrrrrr}
     \toprule
         & \multicolumn{3}{c}{Prop. Corr .Quintil} &     & \multicolumn{3}{c}{Prop. corr./neighbor. Q.} &     & \multicolumn{1}{c}{Bias } &     &     & \multicolumn{1}{c}{MSE} &     &  \\
\cmidrule{2-4}\cmidrule{6-8}\cmidrule{10-10}\cmidrule{13-13}         & \multicolumn{1}{c}{mean} & \multicolumn{1}{c}{SD} & \multicolumn{1}{c}{median} &     & \multicolumn{1}{c}{mean} & \multicolumn{1}{c}{SD} & \multicolumn{1}{c}{median} &     & \multicolumn{1}{c}{mean} &     &     & \multicolumn{1}{c}{mean} &     &  \\
     \multicolumn{15}{l}{\textit{\textbf{univariate prediction}}} \\
     \midrule
     \multicolumn{15}{l}{\textit{Extent of random field (dimensionality in $m^2$)}} \\
     \midrule
     8000 & 0.34 & 0.28 & 0.28 &     & 0.76 & 0.27 & 0.87 &     & -0.040 &     &     & 3.806 &     &  \\
     10000 & 0.45 & 0.33 & 0.44 &     & 0.82 & 0.28 & 0.97 &     & 0.067 &     &     & 3.346 &     &  \\
     15000 & 0.33 & 0.29 & 0.28 &     & 0.75 & 0.30 & 0.88 &     & 0.110 &     &     & 3.664 &     &  \\
     \midrule
     \multicolumn{15}{l}{\textit{(true) Nugget effect}} \\
     \midrule
     0   & 0.41 & 0.30 & 0.39 &     & 0.82 & 0.25 & 0.95 &     & 0.043 &     &     & 2.557 &     &  \\
     0.2 & 0.33 & 0.30 & 0.26 &     & 0.73 & 0.31 & 0.87 &     & 0.048 &     &     & 4.653 &     &  \\
     \midrule
     \multicolumn{15}{l}{\textit{(true) Correlation between variables}} \\
     \midrule
     0.1 & 0.35 & 0.26 & 0.33 &     & 0.78 & 0.27 & 0.91 &     & 0.035 &     &     & 2.716 &     &  \\
     0.5 & 0.37 & 0.31 & 0.31 &     & 0.77 & 0.29 & 0.91 &     & 0.047 &     &     & 3.600 &     &  \\
     0.9 & 0.40 & 0.34 & 0.33 &     & 0.77 & 0.30 & 0.93 &     & 0.055 &     &     & 4.499 &     &  \\
         &     &     &     &     &     &     &     &     &     &     &     &     &     &  \\
     \multicolumn{13}{l}{\textit{\textbf{multivariate prediction (using co-variograms)}}} &     &  \\
    
     \midrule
     \multicolumn{15}{l}{\textit{Extent of random field (dimensionality in $m^2$)}} \\
     \midrule
     8000 & 0.39 & 0.28 & 0.34 &     & 0.77 & 0.24 & 0.86 &     & -0.060 &     &     & 3.449 &     &  \\
     10000 & 0.47 & 0.32 & 0.45 &     & 0.83 & 0.26 & 0.96 &     & 0.045 &     &     & 3.030 &     &  \\
     15000 & 0.38 & 0.28 & 0.35 &     & 0.76 & 0.27 & 0.87 &     & 0.107 &     &     & 3.373 &     &  \\
     \midrule
     \multicolumn{15}{l}{\textit{(true) nugget effect}} \\
     \midrule
     0   & 0.45 & 0.29 & 0.42 &     & 0.83 & 0.22 & 0.93 &     & 0.028 &     &     & 2.233 &     &  \\
     0.2 & 0.37 & 0.30 & 0.33 &     & 0.74 & 0.29 & 0.85 &     & 0.034 &     &     & 4.335 &     &  \\
     \midrule
     \multicolumn{15}{l}{\textit{(true) Correlation between variables}} \\
     \midrule
     0.1 & 0.36 & 0.25 & 0.34 &     & 0.78 & 0.26 & 0.90 &     & 0.033 &     &     & 2.703 &     &  \\
     0.5 & 0.41 & 0.29 & 0.37 &     & 0.78 & 0.26 & 0.89 &     & 0.033 &     &     & 3.361 &     &  \\
     0.9 & 0.47 & 0.32 & 0.40 &     & 0.79 & 0.26 & 0.91 &     & 0.028 &     &     & 3.788 &     &  \\
     \bottomrule
     \end{tabular}%
    \caption*{\footnotesize{Note: Values are summarized over the 200 testpoints; SD: standard deviation; prop.corr. Q.: proportion of predicted values in  correct quintile; corr./neighbor. Q.; proportion of predicted values in  correct or neighbouring quintile; Bias: mean of predicted minus true value in SD units of true outcome; SE: mean standard error of predicted values; MSE: mean squared error of predicted values; No. of points: number of observed points used for local kriging; Correlations: pairwise correlation of the 3 variables in the simulated random field; Density: density of sampled observed points}}
  \label{tab:multi1}
 \end{table}%

 \begin{table}[h!tbp]
   \centering
   \caption{Quality of prediction in different kriging scenarios: multivariate simulation with 3 variables observed at the same locations (collocated. 1.000 known points)}  
      \begin{tabular}{rrrrrrrrrrrrrrr}
     \toprule
         & \multicolumn{3}{c}{Prop. Corr .Quintil} &     & \multicolumn{3}{c}{Prop. corr./neighbor. Q.} &     & \multicolumn{1}{c}{Bias } &     &     & \multicolumn{1}{c}{MSE} &     &  \\
\cmidrule{2-4}\cmidrule{6-8}\cmidrule{10-10}\cmidrule{13-13}         & \multicolumn{1}{c}{mean} & \multicolumn{1}{c}{SD} & \multicolumn{1}{c}{median} &     & \multicolumn{1}{c}{mean} & \multicolumn{1}{c}{SD} & \multicolumn{1}{c}{median} &     & \multicolumn{1}{c}{mean} &     &     & \multicolumn{1}{c}{mean} &     &  \\
     \multicolumn{15}{l}{\textit{\textbf{univariate prediction}}} \\
     \midrule
     \multicolumn{15}{l}{\textit{Extent of random field (dimensionality in $m^2$)}} \\
     \midrule
     8000 & 0.29 & 0.20 & 0.26 &     & 0.67 & 0.25 & 0.72 &     & -0.016 &     &     & 5.077 &     &  \\
     10000 & 0.30 & 0.18 & 0.30 &     & 0.69 & 0.25 & 0.75 &     & 0.106 &     &     & 5.768 &     &  \\
     15000 & 0.29 & 0.20 & 0.27 &     & 0.67 & 0.26 & 0.73 &     & 0.108 &     &     & 4.874 &     &  \\
     \midrule
     \multicolumn{15}{l}{\textit{Density (expected no. of known points)}} \\
     \midrule
     2   & 0.29 & 0.20 & 0.27 &     & 0.68 & 0.25 & 0.73 &     & 0.066 &     &     & 5.239 &     &  \\
     4   & 0.29 & 0.20 & 0.27 &     & 0.68 & 0.25 & 0.73 &     & 0.066 &     &     & 5.238 &     &  \\
     7   & 0.29 & 0.20 & 0.28 &     & 0.68 & 0.25 & 0.74 &     & 0.067 &     &     & 5.242 &     &  \\
     \midrule
     \multicolumn{15}{l}{\textit{(true) Nugget effect}} \\
     \midrule
     0   & 0.32 & 0.20 & 0.31 &     & 0.72 & 0.24 & 0.78 &     & 0.066 &     &     & 4.140 &     &  \\
     0.2 & 0.26 & 0.19 & 0.25 &     & 0.64 & 0.26 & 0.68 &     & 0.066 &     &     & 6.340 &     &  \\
     \midrule
     \multicolumn{15}{l}{\textit{(true) Correlation between variables}} \\
     \midrule
     0.1 & 0.28 & 0.19 & 0.27 &     & 0.71 & 0.25 & 0.78 &     & 0.052 &     &     & 3.570 &     &  \\
     0.5 & 0.29 & 0.19 & 0.28 &     & 0.67 & 0.25 & 0.73 &     & 0.066 &     &     & 5.237 &     &  \\
     0.9 & 0.31 & 0.21 & 0.28 &     & 0.65 & 0.25 & 0.70 &     & 0.080 &     &     & 6.912 &     &  \\
         &     &     &     &     &     &     &     &     &     &     &     &     &     &  \\
     \multicolumn{13}{l}{\textit{\textbf{multivariate prediction (using co-variograms)}}} &     &  \\
     \midrule
     \multicolumn{15}{l}{\textit{Extent of random field (dimensionality in $m^2$)}} \\
     \midrule
     8000 & 0.29 & 0.21 & 0.26 &     & 0.67 & 0.26 & 0.72 &     & -0.014 &     &     & 5.070 &     &  \\
     10000 & 0.29 & 0.19 & 0.29 &     & 0.69 & 0.25 & 0.75 &     & 0.107 &     &     & 5.772 &     &  \\
     15000 & 0.28 & 0.20 & 0.26 &     & 0.67 & 0.26 & 0.73 &     & 0.108 &     &     & 4.865 &     &  \\
     \midrule
     \multicolumn{15}{l}{\textit{Density (expected no. of known points)}} \\
     \midrule
     2   & 0.29 & 0.20 & 0.27 &     & 0.68 & 0.26 & 0.74 &     & 0.067 &     &     & 5.235 &     &  \\
     4   & 0.29 & 0.20 & 0.27 &     & 0.68 & 0.26 & 0.74 &     & 0.066 &     &     & 5.234 &     &  \\
     7   & 0.29 & 0.20 & 0.27 &     & 0.68 & 0.26 & 0.74 &     & 0.067 &     &     & 5.238 &     &  \\
     \midrule
     \multicolumn{15}{l}{\textit{(true) nugget effect}} \\
     \midrule
     0   & 0.32 & 0.20 & 0.31 &     & 0.72 & 0.24 & 0.78 &     & 0.066 &     &     & 4.140 &     &  \\
     0.2 & 0.26 & 0.19 & 0.23 &     & 0.64 & 0.26 & 0.68 &     & 0.068 &     &     & 6.331 &     &  \\
     \midrule
     \multicolumn{15}{l}{\textit{(true) Correlation between variables}} \\
     \midrule
     0.1 & 0.27 & 0.20 & 0.25 &     & 0.71 & 0.26 & 0.79 &     & 0.053 &     &     & 3.558 &     &  \\
     0.5 & 0.28 & 0.19 & 0.27 &     & 0.67 & 0.25 & 0.73 &     & 0.067 &     &     & 5.236 &     &  \\
     0.9 & 0.31 & 0.21 & 0.28 &     & 0.65 & 0.25 & 0.70 &     & 0.080 &     &     & 6.913 &     &  \\
     \bottomrule
     \end{tabular}%
    \caption*{\footnotesize{Note: Values are summarized over the 200 testpoints; SD: standard deviation; prop.corr. Q.: proportion of predicted values in  correct quintile; corr./neighbor. Q.; proportion of predicted values in  correct or neighbouring quintile; Bias: mean of predicted minus true value in SD units of true outcome; SE: mean standard error of predicted values; MSE: mean squared error of predicted values; No. of points: number of observed points used for local kriging ; Correlations: pairwise correlation of the 3 variables in the simulated random field; Density: density of sampled points}}
  \label{tab:multi2}
 \end{table}%

\begin{landscape}
 \begin{table}[htp]
  \centering
      \caption{Quality of prediction for RECORD data with varying the number of known points for kriging, estimated semi-variograms based on all available points }
\resizebox{22cm}{!}{
\begin{tabular}{llrrrrrrrrrrrrr}
\toprule
& & \multicolumn{2}{c}{prop. correct quintile} & & \multicolumn{2}{c}{prop. corr. or neighboring q.} & & \multicolumn{7}{c}{residuum kriging value} \\
\cmidrule{3-4}\cmidrule{6-7}\cmidrule{9-15} \multirow{2}[2]{*}{no of points} & & \multicolumn{1}{r}{\multirow{2}[2]{*}{univariate}} & \multicolumn{1}{r}{\multirow{2}[2]{*}{multivariate}} & & \multicolumn{1}{r}{\multirow{2}[2]{*}{univariate}} & \multicolumn{1}{r}{\multirow{2}[2]{*}{multivariate}} & & \multicolumn{3}{c}{univariate} & & \multicolumn{3}{c}{multivariate} \\
& & & & & & & & \multicolumn{1}{c}{M} & \multicolumn{1}{c}{SD} & \multicolumn{1}{c}{median} & & \multicolumn{1}{c}{M} & \multicolumn{1}{c}{SD} & \multicolumn{1}{c}{median} \\
\cmidrule{1-1}\cmidrule{3-4}\cmidrule{6-7}\cmidrule{9-11}\cmidrule{13-15} 500 & & 0.38 & 0.39 & & 0.88 & 0.86 & & -0.020 & 0.449 & -0.022 & & -0.021 & 0.447 & -0.019 \\
1000 & & 0.43 & 0.42 & & 0.87 & 0.87 & & -0.007 & 0.427 & -0.018 & & -0.008 & 0.426 & -0.022 \\
2000 & & 0.54 & 0.56 & & 0.89 & 0.90 & & -0.024 & 0.357 & -0.010 & & -0.025 & 0.355 & -0.011 \\
5000 & & 0.62 & 0.62 & & 0.94 & 0.94 & & -0.018 & 0.297 & -0.003 & & -0.019 & 0.294 & -0.006 \\
7090 & & 0.66 & 0.66 & & 0.95 & 0.96 & & -0.005 & 0.264 & -0.001 & & -0.006 & 0.263 & -0.001 \\
\bottomrule
\end{tabular}}%
    \label{tab:record1}%
   \caption*{\footnotesize{Note: For every combination of varied parameters, values at the 200 test locations were estimated independently for the 3 values (univariate) and taking into account the co-variograms additionally (multivariate). The 50 nearest points were used for local kriging. Values were summarized over the 200 locations per parameter combination and distributional parameters are given for the proportions of correctly predicted quintiles and for the distributions of estimation errors.\\ no. of points: number of observed points sampled for kriging; prop. correct quintile: proportion of predicted values in the correct quintile; prop. corr./neighb.Q.: proportion predicted values in correct or neighbouring quintile; SD: standard deviation. }  }
 \end{table}

 \end{landscape}
 
\newpage
\begin{landscape}

 \begin{table}[htbp]
   \centering
    \caption{Quality of prediction for RECORD data with varying the number of known points for kriging, estimated semi-variograms based on sampled points only}
     \resizebox{22cm}{!}{
\begin{tabular}{llrrrrrrrrrrrrr}
\toprule
& & \multicolumn{2}{c}{proportion correct quintile} & & \multicolumn{2}{c}{prop. corr. or neighboring q.} & & \multicolumn{7}{c}{residuum kriging value} \\
\cmidrule{3-15} \multirow{2}[2]{*}{no of points} & & \multicolumn{1}{r}{\multirow{2}[2]{*}{univariate}} & \multicolumn{1}{r}{\multirow{2}[2]{*}{multivariate}} & & \multicolumn{1}{r}{\multirow{2}[2]{*}{univariate}} & \multicolumn{1}{r}{\multirow{2}[2]{*}{multivariate}} & & \multicolumn{3}{c}{univariate} & & \multicolumn{3}{c}{multivariate} \\
& & & & & & & & \multicolumn{1}{c}{mean} & \multicolumn{1}{c}{SD} & \multicolumn{1}{c}{median} & & \multicolumn{1}{c}{mean} & \multicolumn{1}{c}{SD} & \multicolumn{1}{c}{median} \\
\cmidrule{1-1}\cmidrule{3-4}\cmidrule{6-7}\cmidrule{9-11}\cmidrule{13-15} 500 & & 0.38 & 0.30 & & 0.86 & 0.79 & & -0.023 & 0.449 & -0.006 & & -0.007 & 0.503 & -0.013 \\
1000 & & 0.44 & 0.39 & & 0.88 & 0.86 & & 0.003 & 0.413 & 0.000 & & 0.001 & 0.439 & -0.001 \\
2000 & & 0.51 & 0.47 & & 0.91 & 0.92 & & -0.008 & 0.371 & -0.003 & & -0.013 & 0.377 & -0.011 \\
5000 & & 0.60 & 0.61 & & 0.94 & 0.94 & & -0.006 & 0.297 & -0.004 & & -0.007 & 0.300 & -0.004 \\
7090 & & 0.66 & 0.66 & & 0.95 & 0.95 & & -0.005 & 0.264 & -0.001 & & -0.007 & 0.263 & -0.001 \\
\bottomrule
\end{tabular} }%
   \label{tab:record2}%
   \caption*{\footnotesize{Note: For every combination of varied parameters, values at the 200 test locations were estimated independently for the 3 values (univariate) and taking into account the co-variograms additionally (multivariate). The 50 nearest points were used for local kriging. Values were summarized over the 200 locations per parameter combination and distributional parameters are given for the proportions of correctly predicted quintiles and for the distributions of estimation errors.\\no. of points: number of observed points sampled for kriging; prop. correct quintile: proportion of predicted values in the correct quintile; prop. corr./neighb.Q.: proportion predicted values in correct or neighbouring quintile; SD: standard deviation. }  }
 \end{table}%

\end{landscape}

\newpage
\begin{landscape}

\begin{table}[htbp]
   \centering
   \caption{Estimated exponential semi-variogram parameters for the 3 neighborhood variables in the RECORD example, varying the  number of sampled known points and the maximum semi-variogram distance }\label{tab.semi.var.Record}
   \resizebox{22cm}{!}{
     \begin{tabular}{crcrrrrrrrrrrrr}
     \toprule
     \multicolumn{1}{l}{\multirow{2}[3]{*}{no. sampled points}} & \multicolumn{1}{r}{\multirow{2}[3]{*}{distance (radius)}} & \multicolumn{1}{l}{Points within } &     & \multicolumn{3}{c}{Deterioration Physical Environment} &     & \multicolumn{3}{c}{Deterioration Social Environment} &     & \multicolumn{3}{c}{Insecurity by others} \\
\cmidrule{5-7}\cmidrule{9-11}\cmidrule{13-15}         &     & \multicolumn{1}{l}{radius: M (SD) } &     & \multicolumn{1}{l}{exp. param.} & \multicolumn{1}{l}{scale par.} & \multicolumn{1}{l}{nugget} &     & \multicolumn{1}{l}{exp. param.} & \multicolumn{1}{l}{scale par.} & \multicolumn{1}{l}{nugget} &     & \multicolumn{1}{l}{exp. param.} & \multicolumn{1}{l}{scale par.} & \multicolumn{1}{l}{nugget} \\
   \multirow{5}[1]{*}{500} & 250 & 0.3 (0.6) &     & 0.491 & 601.194 & 0.116 &     & 0.361 & 2711.814 & 0.119 &     & 0.19 & 356.795 & 0.018 \\
         & 500 & 1.5 (1.5) &     & 0.523 & 324.204 & 0   &     & 0.266 & 761.802 & 0.032 &     & 0.242 & 471.256 & 0 \\
         & 756 & 3.1 (2.6) &     & 0.526 & 358.493 & 0.03 &     & 0.286 & 356.141 & 0   &     & 0.245 & 373.204 & 0 \\
         & 1000 & 5.1 (3.8) &     & 0.504 & 246.043 & 0   &     & 0.227 & 336.397 & 0   &     & 0.22 & 441.175 & 0.007 \\
         & 1250 & 7.4 (5.4) &     & 0.352 & 298.134 & 0.121 &     & 0.203 & 61.398 & 0   &     & 0.194 & 162.664 & 0 \\
     \midrule
     \multirow{5}[2]{*}{1000} & 250 & 0.8 (1.0) &     & 0.477 & 226.293 & 0   &     & 0.234 & 245.975 & 0   &     & 0.21 & 292.23 & 0 \\
         & 500 & 2.9 (2.4) &     & 0.482 & 222.058 & 0   &     & 0.268 & 630.808 & 0.05 &     & 0.223 & 379.763 & 0.015 \\
         & 756 & 6.1 (4.6) &     & 0.523 & 287.227 & 0   &     & 0.243 & 254.983 & 0.002 &     & 0.184 & 283.157 & 0.018 \\
         & 1000 & 10.4 (7.4) &     & 0.485 & 235.694 & 0   &     & 0.277 & 321.253 & 0   &     & 0.236 & 288.216 & 0 \\
         & 1250 & 15.4 (10.4) &     & 0.481 & 266.255 & 0   &     & 0.251 & 265.352 & 0.013 &     & 0.205 & 200.481 & 0.021 \\
     \midrule
     \multirow{5}[2]{*}{2000} & 250 & 1.8 (1.9) &     & 0.51 & 244.599 & 0   &     & 0.221 & 181.818 & 0.008 &     & 0.2 & 243.61 & 0.013 \\
         & 500 & 6.4 (5.0) &     & 0.473 & 237.387 & 0   &     & 0.225 & 182.126 & 0   &     & 0.201 & 214.153 & 0.001 \\
         & 756 & 12.8 (9.7) &     & 0.504 & 201.231 & 0.002 &     & 0.231 & 259.972 & 0.017 &     & 0.197 & 308.088 & 0.015 \\
         & 1000 & 20.7 (15.1) &     & 0.507 & 224.805 & 0   &     & 0.219 & 190.033 & 0.012 &     & 0.19 & 256.552 & 0.022 \\
         & 1250 & 30.7 (21.3) &     & 0.501 & 274.617 & 0   &     & 0.224 & 210.997 & 0   &     & 0.201 & 257.217 & 0 \\
     \midrule
     \multirow{5}[2]{*}{5000} & 250 & 4.3 (3.7) &     & 0.488 & 235.509 & 0   &     & 0.234 & 227.92 & 0.008 &     & 0.206 & 241.068 & 0.007 \\
         & 500 & 15.6 (11.2) &     & 0.492 & 247.74 & 0   &     & 0.232 & 203.003 & 0.002 &     & 0.207 & 239.621 & 0.003 \\
         & 756 & 31.5 (22.1) &     & 0.489 & 269.839 & 0   &     & 0.238 & 226.881 & 0   &     & 0.209 & 233.267 & 0 \\
         & 1000 & 51.7 (35.6) &     & 0.493 & 250.548 & 0   &     & 0.233 & 215.678 & 0.001 &     & 0.205 & 269.909 & 0.007 \\
         & 1250 & 76.2 (51.2) &     & 0.477 & 238.951 & 0   &     & 0.222 & 200.809 & 0   &     & 0.2 & 220.586 & 0 \\
     \midrule
     \multirow{5}[2]{*}{7090} & 250 & 6.1 (4.9) &     & 0.489 & 249.71 & 0   &     & 0.233 & 209.682 & 0   &     & 0.204 & 225.641 & 0.001 \\
         & 500 & 22.2 (15.8) &     & 0.489 & 249.71 & 0   &     & 0.233 & 209.682 & 0   &     & 0.204 & 225.641 & 0.001 \\
         & 756 & 45.1 (31.8) &     & 0.489 & 249.71 & 0   &     & 0.233 & 209.682 & 0   &     & 0.204 & 225.641 & 0.001 \\
         & 1000 & 73.9 (51.2) &     & 0.489 & 249.71 & 0   &     & 0.233 & 209.682 & 0   &     & 0.204 & 225.641 & 0.001 \\
         & 1250 & 108.8 (73.7) &     & 0.489 & 249.71 & 0   &     & 0.233 & 209.682 & 0   &     & 0.204 & 225.641 & 0.001 \\
     \bottomrule
  \end{tabular}}%
      \caption*{\footnotesize Note: distance: maximum distance of semi-variograms; exp. param.: estimated parameter of exponential semi-variogram; scale par.: scale  parameter of estimated semi-variogram  with scale = range/3.
      }
    \end{table}%

\end{landscape}

\section*{Figures}
  
\begin{figure} [htp]
\centering
\includegraphics[scale=0.4]{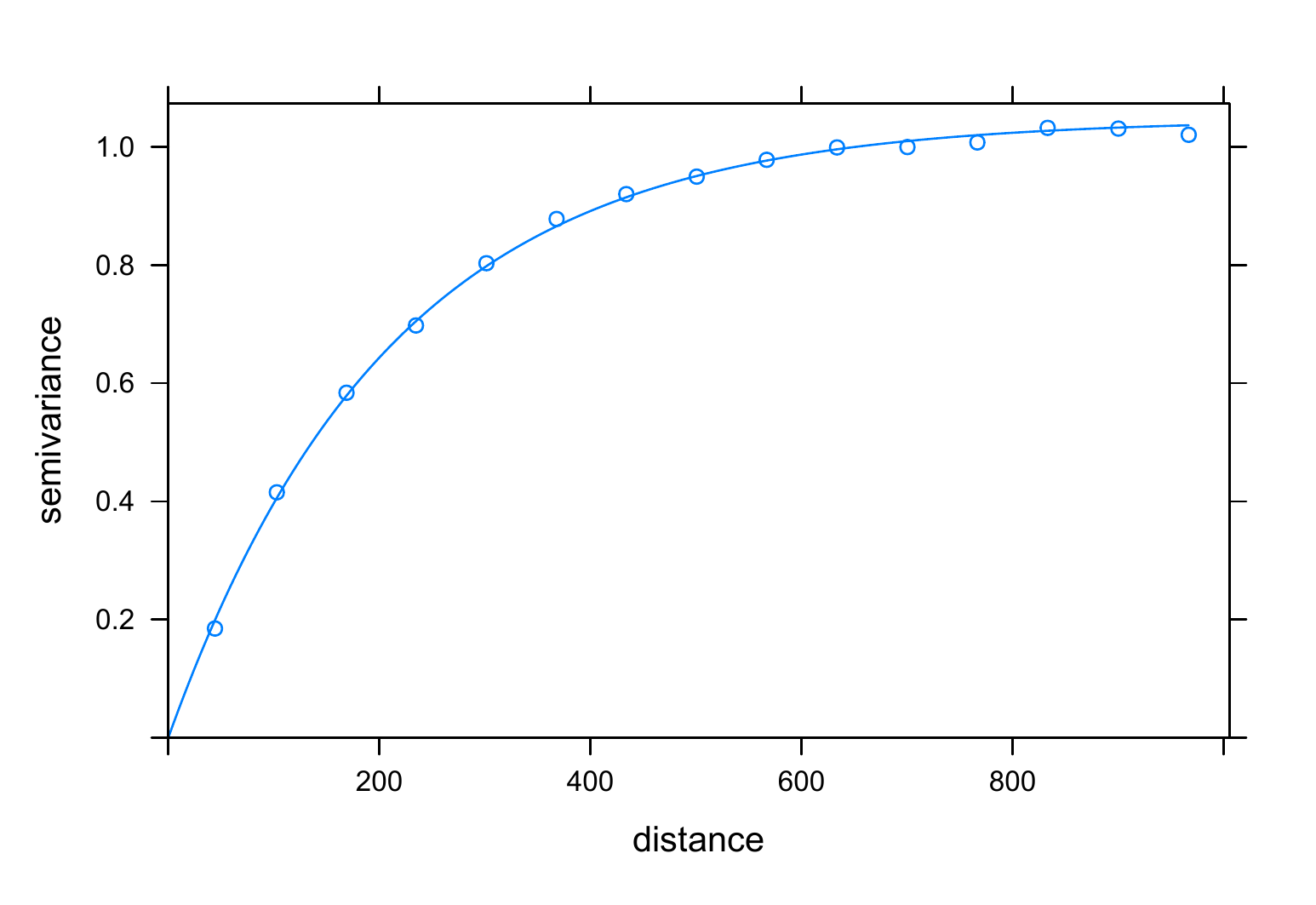}
\caption{Example of an exponential semi-variogram fitted to sampled data (in 1300 sampled points) from a simulated random field of $8000 * 8000m$ with nugget $c_0 = 0$, variance $\sigma^2 = 1$ = partial sill $\sigma^2_0$, practical range = 600} \label{fig:expsemi}
\end{figure}

\end{document}